\documentstyle[11pt,aaspp4,tighten,flushrt]{article}

\lefthead{Anderson \& Statler}
\righthead{Mean Stellar Motions in Elliptical Galaxies}

\begin{document}

\title{Streamlines of the Mean Stellar Motions in Elliptical Galaxies}
\author{R. F. Anderson\altaffilmark{1,3} and 
Thomas S. Statler\altaffilmark{1,2,4}}

\altaffiltext{1}{Department of Physics and Astronomy, University of North
Carolina, Chapel Hill, NC 27559-3255}
\altaffiltext{2}{Present address: Department of Physics and Astronomy, Ohio
University, Athens, OH 45701}
\altaffiltext{3}{boband@physics.unc.edu}
\altaffiltext{4}{tss@coma.phy.ohiou.edu}

\begin{abstract}
The stellar velocity fields of elliptical galaxies hold clues to their 
dynamical structure and origin. The construction of velocity field 
models is greatly simplified by assuming an approximate geometrical form for
the streamlines of the mean stellar motions, for example, representing
the streamlines of short-axis and long-axis tube orbits by coordinate lines
in a confocal coordinate system. A single confocal system precisely fits
the mean motions of all tube orbits in a St\"ackel potential; but these
potentials are not sufficiently general. Here we test the conjecture that
confocal streamlines may still be a valid approximation for
more realistic triaxial systems. We numerically integrate orbits
in Schwarzschild's \markcite{Sch93}(1993) logarithmic potential. Six sets
of axis ratios are used; in each, $\sim 50$ orbits, comprising short-axis
and long-axis tubes as well as some resonant families, are
integrated for $\sim 20000$ dynamical times and the average velocity is
found in each of $\sim 4000$ spatial cells. Confocal streamlines are
compared to the velocity field by finding the RMS magnitude of the
cross product between the velocity vectors and the streamlines. Minimizing
this quantity yields a best fit confocal system for each orbit. We find that
the great majority of orbits at a given energy in each potential can be
fitted by nearly identical confocal systems. There are statistically
significant differences between the average streamline parameters obtained
for different orbit families, but the differences are small. We show that
the fitted parameters reproduce, to high accuracy, the location of the
boundary between short axis and outer long axis tubes, which is a direct
measure of the triaxiality of the potential. These results strongly support
efforts to obtain accurate statistical measurements of triaxiality from
kinematic observations and reasonably simple velocity field models.
\end{abstract}


\section{Introduction}

The stellar velocity fields of elliptical galaxies are important 
diagnostics of their intrinsic shapes and dynamical structure, which in
turn hold clues to their origin and evolution. Methods to extract
structural information from the observable velocity fields, initially
developed by Binney \markcite{Bin85}(1985), have been greatly expanded
upon by Franx, Illingworth and de Zeeuw \markcite{Fra91}(1991), Tenjes
{\em et al.\/} (\markcite{Ten92}1992, \markcite{Ten93}1993), and
Statler (\markcite{Sta94a}1994a,\markcite{Sta94b}b), among others. The
construction of velocity field models is a complex task, particularly if
the dynamical model is required to be fully self-consistent; some velocity
field models of this sort have been laboriously calculated ({\em e.g,\/}
Statler \markcite{Sta91}1991, Arnold {\em et al.\/} \markcite{Arn94}1994).
The methods cited above, however, sidestep the computational bottleneck
of self-consistency by adopting geometrical approximations for the
streamlines of the mean stellar motions and then solving the equation of
continuity.

In most of these models, the mean stellar flow is assumed to follow
elliptical streamlines in parallel planes. In Binney's \markcite{Bin85}(1985)
models, the streamlines are tangent to the equidensity surfaces and
perpendicular to the long or short axes of the figure. The models of Tenjes
{\em et al.\/} \markcite{Ten92}(1992) are similar but allow an arbitrary
tilt with respect to the principal axes. Franx {\em et al.\/}
\markcite{Fra91}(1991) use circular or elliptical streamlines aligned
either with or contrary to the equidensity surfaces and perpendicular
to the short axis. (Cf. Figure 1 of Statler \markcite{Sta96fv}1996.)

A somewhat different approach is taken by Statler \markcite{Sta94a}(1994a),
who calculates mean stellar motions as vector sums of two intersecting
collisionless flows created by short-axis and long-axis tube orbits.
The streamlines for these flows are assumed to follow coordinate lines
in a confocal ellipsoidal coordinate system. In the special class of
St\"ackel potentials, a single such confocal system gives a set of
streamlines that precisely fit the mean motions of all orbits in a given
potential; consequently the same system fits the total short-axis
and long-axis mean flows, regardless of the form of the distribution
function. Unfortunately, the Stackel potentials are not sufficiently general
to be realistic models for elliptical galaxies (Binney \markcite{Bin87}1987,
Merritt \& Fridman \markcite{MeF96}1996). In non-St\"ackel
potentials, one should not expect a set of confocal streamlines to
precisely fit the mean motion of an arbitrary orbit, or that streamlines
fitted to different orbits will come from the same confocal system.
Nonetheless, it is possible that confocal streamlines may be a valid
approximation for the combined mean motions of many orbits in realistic
triaxial systems.

In this paper we test this conjecture, by fitting confocal streamlines
to the mean velocities of individual numerically integrated tube orbits in 
a set of realistic non-rotating scale-free potentials (Schwarzschild
\markcite{Sch93}1993). For each orbit, mean velocity vectors are computed
over a fine spatial grid; we then find the parameters of the confocal
system for which the relevant coordinate lines are most nearly tangent to
the mean velocities. For all axis ratios tested, the parameters of the
fitted confocal system are weak functions of the integrals of motion.
Thus, at a given energy in a given potential, a single confocal system
provides an excellent fit to the mean motions of short-axis tubes, long-axis
tubes, and the dominant higher-order resonances. Scaling
to other energies is straightforward since the potentials are scale-free.
We conclude that the assumption of confocal streamlines can be used
to create realistic velocity field models for elliptical galaxies that do
not have significant figure rotation.
 
The remainder of this paper is arranged as follows. In section 2 we
remind the reader of the properties of the Schwarzschild \markcite{Sch93}(1993)
logarithmic potential and describe our methods for numerical
integration of the orbits and fitting of streamlines. In section 3 we
present our results and in section 4 discuss their implications.

\section{Approach}
\subsection{Model Potentials}

We adopt for this study Schwarzschild's \markcite{Sch93}(1993; hereafter S93)
version of the scale-free logarithmic potential, given by
\begin {eqnarray}
\label{e:logpotential}
V & = & {1\over2}\Biggl(\ln r^2+c_1{z^2\over r^2}+c_2{y^2\over r^2}+
c_3{z^4\over r^4}+c_4{z^2y^2\over r^4} \nonumber \\
  &   & \phantom{{1\over2}\Biggl(\ln r^2}
+c_5{y^4\over r^4}+c_6{z^6\over r^6}+c_7{z^4y^2\over r^6}+ 
c_8{z^2y^4\over r^6}+c_9{y^6\over r^6}\Biggr).
\end{eqnarray}
The constants $c_1$ through $c_9$ depend on the axis ratios of the mass
density distribution, which is close to, though not exactly, ellipsoidal.
These constants are tabulated by Schwarzschild for six sets
of axis ratios, which we also adopt here. In columns 2 and 3 of
Table \ref{t:models} we list for each figure the triaxiality
$T_m = (1-b_m^2)/(1-c_m^2)$ and the short-to-long axis ratio
$c_m$ of the mass distribution (where, in the definition of $T_m$,
$b_m$ is the middle-to-long axis ratio.)
We can obtain analogous quantities for the potential, despite the slightly
non-ellipsoidal shape of the equipotential surfaces, from the 
points at which these surfaces cross the
principal axes. If these crossings occur at locations $x_c$, $y_c$, and
$z_c$ along the $x$, $y$ and $z$ axes, then
\begin{equation}
\ln x_c^{2}
= \ln y_c^{2}+c_{2}+c_{5}+c_{9}
= \ln z_c^{2}+c_{1}+c_{3}+c_{6}.
\end{equation}
We thus obtain
\begin{equation}
b_p = {y_c \over x_c} = e^{-{c_2+c_5+c_9 \over 2}}, \qquad
c_p = {z_c \over x_c} = e^{-{c_1+c_3+c_6 \over 2}}.
\end{equation}
The triaxiality $T_p$ and axis ratio $c_p$ of the potential are given in
columns 4 and 5 of Table \ref{t:models}.

In this paper we eschew the simpler version of the logarithmic potential,
$V = \onehalf \ln (x^2 + y^2/b^2 + z^2/c^2)$. This version is less
astrophysically realistic than equation (\ref{e:logpotential})
because it derives from a density that becomes dimpled on the $z$
axis as the flattening increases and turns negative at $c < 0.707$
(Binney \& Tremaine \markcite{BandT}1987). We have, however, repeated our
calculations for the analogue of Model 2, and obtain
results qualitatively the same as those described below.

The logarithmic potential (\ref{e:logpotential})
is unlike most triaxial St\"ackel potentials
in two important ways. First, the potential can remain significantly
nonspherical to large radii; second, both the potential and the density
are singular at $r=0$. The inner $r^{-2}$ density profile implied by the
logarithmic potential is consistent with the steeper surface
brightness cusps observed by the HST WF/PC (Lauer {\em et al.\/}
\markcite{Lau95}1995). The triaxial logarithmic potential is the best
studied case in which a central mass concentration causes the majority of
regular box orbits to be replaced by isolated resonant islands in a
stochastic sea (Miralda-Escud\'e \& Schwarzschild \markcite{Mir89}1989,
Lees \& Schwarzschild \markcite{Lee92}1992); it is conjectured (Merritt
\& Fridman \markcite{MeF96}1996, Merritt \& Valluri \markcite{MeV96}1996)
that this loss of box orbits may render triaxial equilibria impossible for the
observed steep-cusped systems, or at least cause them to evolve on a time
scale comparable to or less than a Hubble time. We take the view that
determining whether or not steep-cusped ellipticals actually are triaxial,
and consequently whether or not such a mechanism is at work, is fundamentally
an observational problem. Our broad goal here is to have the
ability to construct triaxial models that are in principle falsifiable by
kinematic observations.

\subsection{Numerical Integration of Orbits}

Since the potential is scale-free, the orbital structure is the same at
all energies. Following standard practice, we calculate orbits at $E=0$.
This choice plus the overall constant of proportionality in equation
(\ref{e:logpotential}) set the units of length and time. The zero-energy
$x$-axial orbit (which is unstable for all cases studied here) has unit
amplitude and period $T_x = 2(2\pi)^{1/2} \approx 5$ for all axis ratios;
the circular orbit in the spherical potential lies at
$r_c = e^{-1/2} \approx 0.6$ and has period $T_c = 2 \pi e^{-1/2} \approx 3.8$.

A test particle launched from a given point in the $x$-$z$ plane has
a velocity whose magnitude is fixed by the potential at that point and the
total energy (here zero). If the direction is taken to be perpendicular
to the $x$-$z$ plane, then its orbit is determined uniquely by the
initial point $(x_0,z_0)$. This defines the ``$x$-$z$ start space'', in
which \markcite{Sch93}S93 maps the boundaries of the
major orbit families. The $x$-$z$ start space is not complete, since not
all orbits cross the $x$-$z$ plane with $v_x = v_z = 0$. It does, however,
include all of the short-axis (S) tubes and the inner (I) and outer (O)
long-axis tubes, which are responsible for the mean rotation in
models with stationary figures. Since the I tubes are low angular momentum
orbits, they generally contribute far less to the observable signature
than the other families except possibly in very prolate figures.
We pick, for each figure, $\sim 50$ orbits from the
O and S tube regions. We include a few I tubes
in the models in which they are potentially important. In models 3 and 4
we also include the resonant ``saucer'' (s) orbits which replace S
tubes in a significant part of the start space. For brevity
we refer to the orbits by model number and start space coordinates;
{\em e.g.,\/} $(2,.63,.12)$ is the orbit launched
from $(x_0,z_0) = (.63,.12)$ in Model 2.

Orbits are integrated using a fourth order Runge-Kutta
method with a variable time step. The step size is set equal to
$0.01 r$, where $r$ is the distance of the particle
from the origin at the start of the step. Energy conservation is always
better than $10^{-6}$ (where typical kinetic energies are of order $10^{-1}$)
over integrations of several tens of thousands of dynamical times. Each
orbit is first integrated to $t=5000$ to determine its family and maximum
extent in the $x$, $y$, and $z$ directions. The volume bounded by the planes
$x=\pm|x_{\rm max}|$, $y=\pm|y_{\rm max}|$, $z=\pm|z_{\rm max}|$
is then divided into cells, with 20 cell widths in each dimension.
Owing to the symmetry of the potential, each $(x_0,z_0)$ in the first
quadrant represents four orbits related by reflection. We enforce symmetry
of the grid across the principal planes to simplify averaging the computed
orbit with its reflections. The number of cells $N_c$ actually visited by
the orbit is typically $\sim 4000$, but depends on eccentricity and varies
from $\sim 1000$ to $\sim 6000$. Because we average reflections of each orbit
the number of {\em independent\/} grid points is one fourth as large.
Once the grid is defined, an integration to $t=80000$ is performed and the
time-average velocity vector calculated for each cell $i$. Note that the mean
velocity is not equal, either in magnitude or direction, to the instantaneous
speed of the particle on any passage though the cell, except in the case of
periodic orbits. These vectors are then normalized to produce a field of
velocity unit vectors $\hat{v}_i$ in three dimensional space.

\subsection{Fitting of Streamlines}

In the confocal ellipsoidal coordinate system
$(\lambda,\mu,\nu)$,\footnote{We use the notation for confocal coordinates
defined by Lynden-Bell \markcite{Lyn62}(1962) and reintroduced by de Zeeuw
\markcite{deZ85}(1985). Coordinate transformations and other basic formulae
can be found in either of those references.}
streamlines for S tubes and saucer orbits should lie close to
lines $\mu = {\rm constant}$, and those for O and I tubes close to lines
$\nu = {\rm constant}$. The shapes of the coordinate lines depend on two
parameters, $(T,\Delta)$, which determine the locations of the foci. In terms
of the conventional notation, $T = (\beta - \alpha) / (\gamma - \alpha)$
and $\Delta = (\gamma-\alpha)^{1/2}.$
The points $z=\pm \Delta$ mark the foci, and the polar angle
$\theta_f= \sin^{-1}T^{1/2}$ the asymptote, of the focal hyperbola, which
in St\"ackel potentials is the boundary between O and S tubes in the
$x$-$z$ start space. In the limit $\Delta \to 0$,
the focal hyperbola collapses onto the asymptote, the coordinate
system becomes scale free, and motion along $\mu = {\rm constant}$ or
$\nu = {\rm constant}$ streamlines is over spherical shells (see
Figures 2, 11, and 12 of Statler \markcite{Sta94a}1994a.) For finite
$\Delta$ the mean motion is over ellipsoidal shells that become progressively
rounder at larger radius.

For each orbit we obtain, at the
center of each cell $i$, the unit vector $\hat{\tau}_i$ in the $\mu$ or $\nu$
direction (depending on the family classification done previously) and resolve
it into cartesian components. We then form the goodness of fit parameter
\begin{equation}
\label{e:qsquared}
Q^2 = {1 \over N_c}\sum_{i=1}^{N_c}
\left|\hat{\tau}_i \times \hat{v}_i\right|^2,
\end{equation}
and minimize with respect to $(T,\Delta)$. The two-dimensional minimization
is straightforward and can be done by standard numerical methods.

\begin{figure}[t]
\epsscale{.7}
\plotone{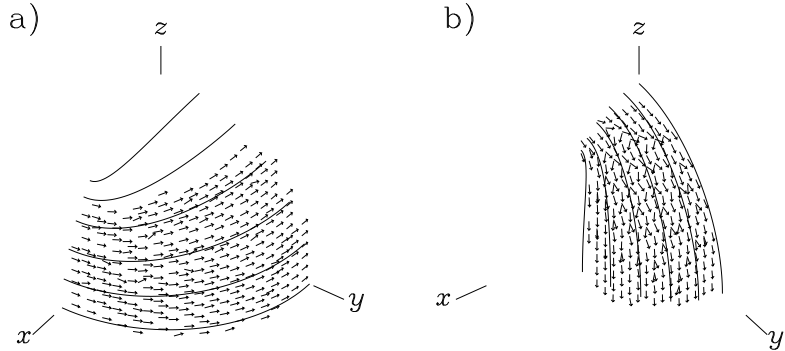}
\caption{\footnotesize
Examples of confocal streamline fits to the mean motions in two thin tube
orbits: (a) the S tube launched from $(x_0,z_0)=(.48,.35)$ in Model 1;
(b) the O tube launched from $(x_0,z_0)=(.23,.44)$ in Model 3. Arrows show
the numerically computed normalized velocity field $\hat{v}$, solid curves
representative streamlines in best fit confocal system. The S tube
illustrates a very good fit, the O tube a fair fit typical of orbits
close to family boundaries.
\label {f:examples}}
\end{figure}

\section{Results}

Two examples of streamline fits are shown in figure \ref{f:examples}.
We specifically show thin tube orbits here for
clarity, so that streamlines on one shell suffice. The S tube
$(1,.48,.35)$ is fitted by the $\mu = {\rm constant}$ lines
on the shell $\lambda+\alpha=0.309$ in a coordinate system with $T=0.481$,
$\Delta=0.263$.  The O tube $(3,.23,.44)$ is fitted by $\nu = {\rm
constant}$ lines on the $\lambda+\alpha=0.101$ shell in the system with
$T=0.551$, $\Delta=0.214$. The quality of the S tube fit is very good
($Q=0.096$), whereas the O tube fit is formally a bit worse ($Q=0.245$)
owing to effects we discuss in section 3.4 below. 

Figure 1 demonstrates that the velocity fields of individual orbits can be
fitted by streamlines built from a confocal coordinate system. But
the more central issue is whether the {\em same\/} coordinate
system can adequately fit a wide variety of orbits in the same potential.

\begin{figure}[p]
\epsscale{.84}
\plotone{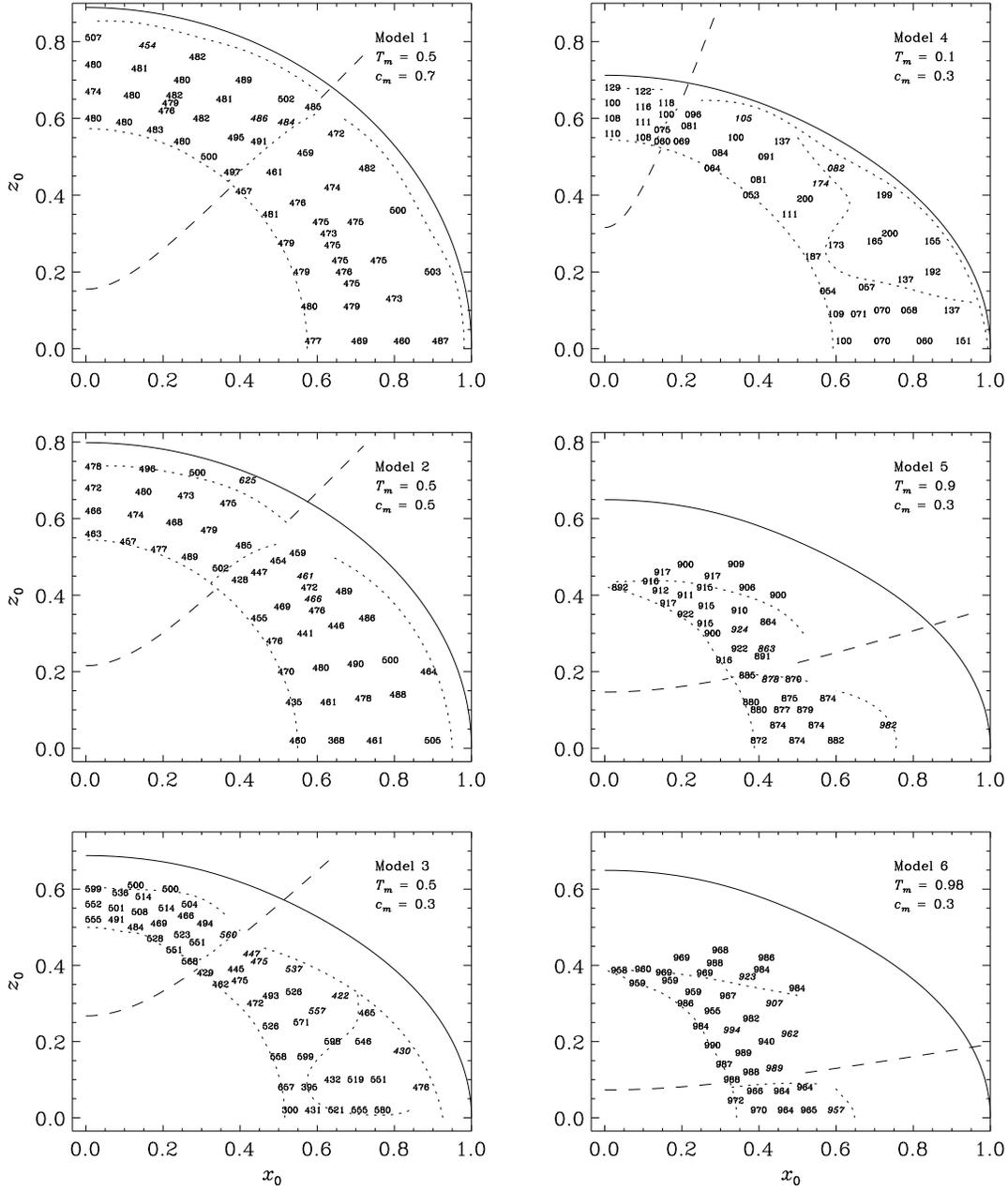}
\caption{\footnotesize
Best-fit streamline triaxiality parameters $T$ are shown in the $x$-$z$
start space for each model (leading decimal points omitted for clarity).
In each panel, circulating orbits lie between the zero-velocity surface
({\em solid line\/}) and the locus of infinitesimally thin O and S tubes
({\em roughly concentric dotted line\/}).
Other {\em dotted lines\/} mark the boundaries between orbit families.
O and S tubes occupy the major regions in the upper left and lower right,
respectively. I tubes lie above the O tubes, and saucers occupy a region
within the S tubes in Models 3 and 4. The small variation in $T$ values
for each model indicates the success of the confocal streamline assumption
for most orbits (section 3.1). {\em Italics\/} indicate orbits identified
with specific resonances (section 3.4). The focal hyperbola defined by
the average $T$ and $\Delta$ for each model ({\em dashed line}) reproduces
the actual O/S boundary with excellent accuracy (section 3.3).
\label {f:T}}
\end{figure}

\subsection{Triaxiality Measured by Streamlines}

Our main result is presented in Figure \ref{f:T}. Each panel in the figure
shows the $x$-$z$ start space for one of the models. The solid curve
marks the zero-velocity surface. The inner, roughly quarter-circular,
dotted line marks the locus of infinitesimally thin S and O tubes.
Launching points interior to this locus produce copies of orbits
launched exterior to it; thus all circulating orbits lie in the annulus
between the zero-velocity surface and the thin-tube boundary. Other
dotted lines in the figure mark the approximate boundaries between the
major orbit families, though their exact locations are 
problematical because of the abundance of high-order resonant islands
and stochastic layers near these boundary regions. (Ignore, for the
moment, the long-dashed hyperbola, which we discuss in section 3.4.)
The O and S tubes are
the major areas marked off in the upper left and lower right,
respectively. Box orbits lie outside the S tubes. I tubes lie above
the O tubes, but do not extend all the way to the zero-velocity surface
because the thin I tube locus intervenes. In the panels for Models 3 and
4 we also show the saucer orbit region within the S tubes. The reader
is referred to \markcite{Sch93}S93 for further details
of the phase space structure.

At the location of each computed regular orbit we indicate the fitted
value of the streamline triaxiality parameter $T$ (omitting the leading
decimal point for clarity). Three things are evident from the figure:
First, the variation of $T$ over each start space is small; in no case
is the standard deviation greater than $0.06$. Second, there is little
indication of systematic trends across the start spaces, except at
the major family boundaries in some of the models. Finally, the fitted $T$
values are quite close to the true triaxialities $T_m$.

These results are quantified in columns 2 and 3 of Table \ref{t:averages},
where we give means and standard deviations of $T$ obtained for the major
orbit families separately and for all orbits taken together. (These are
unweighted means; we make no attempt to compute exact phase space volumes
for the orbits or to correct for our non-uniform sampling of the start
spaces.) There are differences in $\langle T \rangle$ that are significant
at the $3\sigma$ or greater level between the O and S tubes in Models 1
and 5, and between the S tubes and saucers in Model 4. The O and S tubes
differ by $\sim 2\sigma$ in Models 2 and 4, and by slightly
over $1\sigma$ in Models 3 and 6. The $T$ values of the O and I tubes are
indistinguishable.

The average $\langle T \rangle$ over all orbits is repeated for each model
in the last column of Table \ref{t:models}. There is excellent
agreement between the triaxialities obtained from the streamline fits
and the actual density and potential triaxialities of the models. In
fact, there is a weak tendency for the streamline fits to reproduce $T_p$
more accurately than $T_m$, though the set of mass models tested is too
small to establish this firmly. It may be merely
fortuitous, that the non-monotonic variation of $T_p$ with $c_m$ at
fixed $T_m = 0.5$ is reproduced qualitatively by the streamline fits.

\subsection{Radial Mean Motions}

Radial mean motions are associated with asphericity of the fiducial
shells on which the streamlines lie, and are measured by
the $\Delta$ parameter. The results for $\Delta$ are summarized in
columns 4 and 5 of Table \ref{t:averages}. The standard deviations are, for the
most part, substantially larger than for $T$. Differences in the means
between the O and S tubes are present at the $3\sigma$ level in Models
3 and 5, and at the $2\sigma$ level in models 2 and 6. The S tubes and
saucers in Model 4 also differ at the $2\sigma$ level. The differences
are in the sense that O tubes tend toward higher $\Delta$ and the
saucers toward lower $\Delta$ compared to S tubes.

\begin{figure}[t]
\epsscale{.7}
\plotone{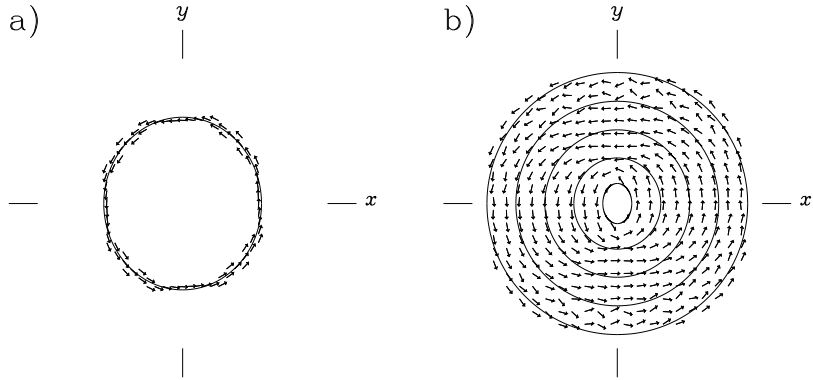}
\caption{\footnotesize
Normalized velocity fields of two nearly planar orbits: (a) (2,.55,.02);
(b) (2,.90,.02). {\em Solid lines\/} show the streamline fits.
Confocal streamlines
correctly reproduce the greater elongation of the fat tube (b)
toward smaller radii, but comparable elongations are reached
at smaller radii compared to the thin tube.
\label {f:elongation}}
\end{figure}

Roughly half the total standard deviation
in $\Delta$ arises from a systematic inward radial gradient in
the start space, reflecting a greater average elongation of the velocity
fields of thin tubes compared with fat tubes. This effect is illustrated
in Figure \ref{f:elongation}, which shows the normalized velocity fields
of the nearly planar orbits (2,.55,.02) and (2,.90,.02) along with
their streamline fits. The confocal fit
correctly reproduces the progressively greater elongation of the fat tube
velocity field toward smaller radius; but comparable elongations are reached
at smaller radii compared to the thin tube. Since the fits average over
volume this results in a smaller $\Delta$ for a fat tube than for a thin
tube at the same energy. This means that adopting the elongation of thin
tubes for all tubes would lead to an overestimate of the radial mean motions.

\subsection{The O Tube/S Tube Boundary}

The averages, $\langle T \rangle$ and $\langle \Delta \rangle$, over all
orbits in a given model suggest an overall best-fit focal hyperbola with
foci at $z=\pm\langle \Delta \rangle$, vertices at $z=\pm\langle \Delta
\rangle(1-\langle T \rangle)^{1/2}$, and asymptotes
at $\sin^{-1}\langle T \rangle^{1/2}$ from the $z$ axis.
These hyperbolae are drawn as
long-dashed curves in Figure \ref{f:T}. The agreement between the fitted
hyperbolae and the actual O tube/S tube boundaries is
remarkable. Remember that the fits are obtained solely from the mean
velocity fields of the orbits, entirely without regard to the orbital
{\em frequencies\/}
which actually determine where the boundaries lie. Given the small
variances in $T$ and $\Delta$, even a small number of orbits quite far
from the boundary would suffice for a very good estimate.

The location of the O tube/S tube boundary is a direct measure of the
triaxiality of the potential. That this quantity can be so accurately
recovered from the (albeit 3-dimensional) mean orbital velocities
compellingly supports the notion that triaxialities of real systems can
be measured from their projected mean velocity fields.

\subsection{Goodness of Fit}

\begin{figure}[p]
\epsscale{.84}
\plotone{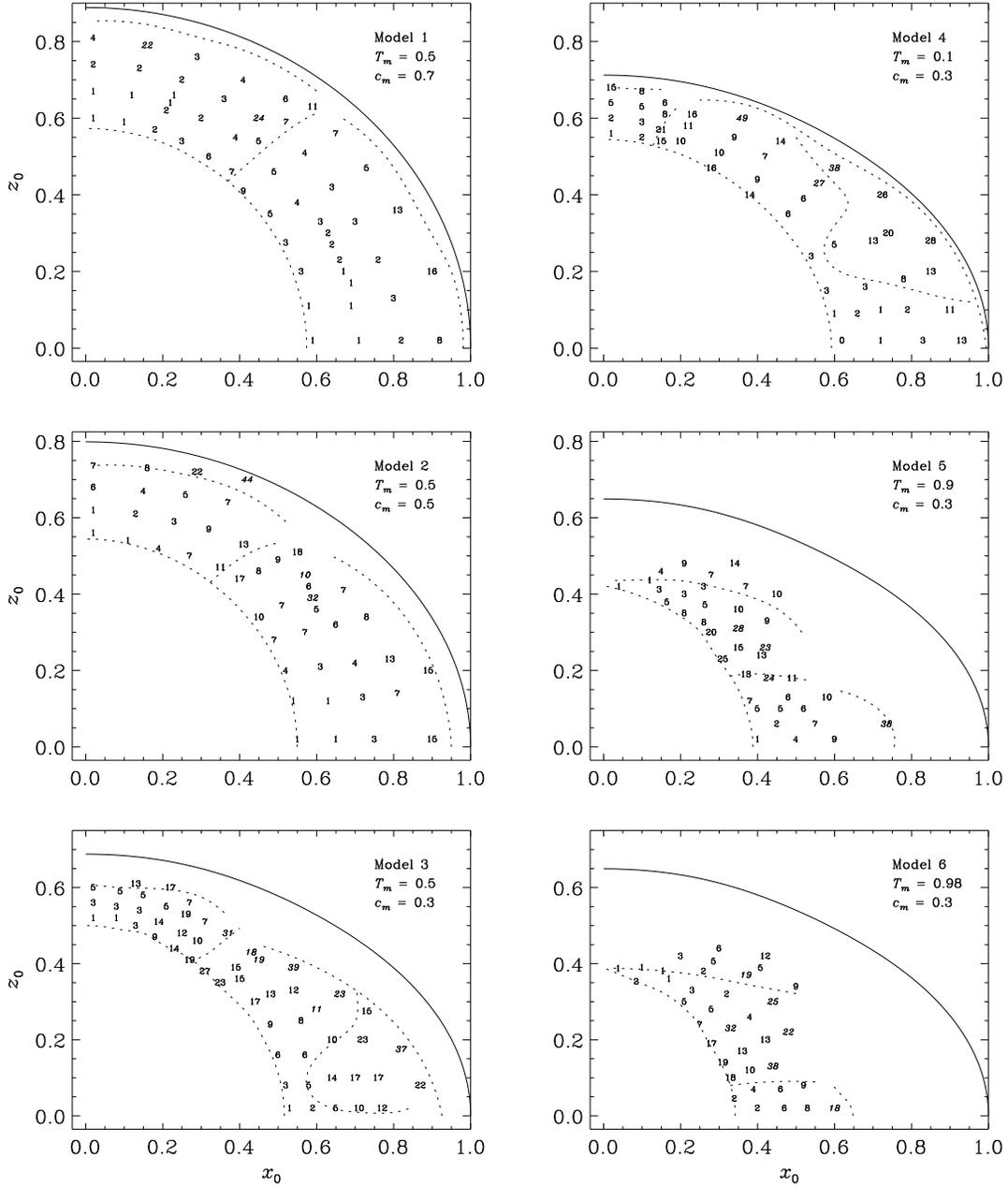}
\caption{\footnotesize
Start spaces as in Figure \protect{\ref{f:T}}, showing the goodness of fit
parameter $\sin^{-1}Q$ in degrees (cf. equation [\ref{e:qsquared}]); this
is essentially the RMS angle between the velocity field and the fitted
streamlines (section 3.4).
\label {f:Q}}
\end{figure}

Figure \ref{f:Q} shows values of $\sin^{-1}Q$ (in degrees) plotted in the
start spaces; this quantity is essentially the RMS angle between the
velocity field and the fitted streamlines. It is under $10\arcdeg$ for just
under two thirds of the computed orbits, and under $23\arcdeg$ for 90\%.
The larger values of $\sin^{-1}Q$ tend to occur close to the O/S
boundary and among the fatter tubes, but also occasionally stand out as
isolated poor fits in otherwise well-behaved regions.

The tendency for poorer fits among the fatter tubes is attributable to
the radial gradient in the velocity field elongation shown in Figure
\ref{f:elongation}b, but the situation near the O/S boundary is more
complicated. In this region the time scale for an even wrapping of the
invariant tori increases, and there is an abundance of high order
resonant islands. Typically the result is a sort of ``braided'' velocity
field, as one can see in Figure \ref{f:examples}b and in the top and bottom of
Figure \ref{f:elongation}b. For orbits trapped by a resonance
this effect will not disappear when the orbit is run for a longer
time or averaged with its reflections. However, this small scale
structure is not a serious problem since it is unlikely to be resolved
in any realistic observation, and can be seen to average out and match
the streamlines in a coarser gridding of the velocity field.

We can identify nearly all of the isolated poor fits with specific high order
resonances or slowly-diffusing regions of the stochastic web. These
orbits are indicated by italics in Figures \ref{f:T} and \ref{f:Q}.
Some resonances of comparatively low order produce substantial radial
mean motions which do not cancel out when reflections are averaged, because
the reflections do not visit the same regions of space. This effect is
not very severe for the 1:1:2 resonance around which the saucer orbits lie,
but does account for the consistently poorer fits compared to the
surrounding S tubes. In general, the higher order the
resonance, the better the fit, owing to the finer scale of velocity field
braiding; good examples are orbits $(2,.59,.39)$ and $(2,.57,.45)$,
apparently associated with 19:19:22 and 85:85:97 resonances, respectively.
Finally, orbit $(4,.36,.60)$ is genuinely stochstic and has a nearly random
velocity field with a very weak residual net circulation about the $x$
axis for the long integration performed.

\section{Discussion}

We have tested the viability of the {\em ansatz\/} of confocal streamlines for
the mean stellar motions in elliptical galaxies by fitting such streamlines
to velocity fields of individual orbits integrated in six scale-free
potentials. We find that the great majority of orbits in each potential can
be fitted by confocal systems with nearly the same $T$ parameter and a
fairly narrow range of $\Delta$ parameters. While there are statistically
significant differences between the average $T$ and $\Delta$ values
obtained for O tubes, S tubes, and saucers, the differences are small enough
that they can almost certainly be neglected in more complete models of galactic
velocity fields. The accuracy with which the mean $T$ and $\Delta$ for
the computed orbits in a given potential recover the location of the
O tube/S tube boundary is impressive, and supports the use of reasonably
simple dynamical models to obtain accurate statistical measurements of
triaxiality from kinematic observations. Forthcoming papers will report
on shape measurements using streamline-based velocity field models
for NGC 1700, NGC 3379, NGC 4472, and the Davies and Birkinshaw (1988)
sample of radio ellipticals. Such measurements may ultimately lead, for
instance, to an understanding of the role of projection effects in
determining the width of the fundamental plane, as well as to
insights into the range of mechanisms resonsible for elliptical galaxy
formation.

It is, of course, a vast oversimplification to expect a single confocal
system to {\it globally\/} fit any non-St\"ackel potential. In this study
we have considered only orbits at zero energy. In scale-free potentials
the orbital structure is self-similar: each orbit at $E=0$ can be scaled
to a copy of itself at energy $E$ according to
$(x_0,z_0) \to (e^E x_0, e^E z_0)$. For this orbit our results would scale
as $\Delta \to e^E \Delta$, $T \to T$, producing a set of similar streamlines.
This is the ``locally confocal'' approximation used by Statler
\markcite{Sta94a}(1994a). Complications arise, however, when similar orbits
overlap, since the elongations of the inner streamlines of the higher-energy
copy will not match those of the outer streamlines of the lower-energy
copy. In practice this will produce some intermediate elongation for the
streamlines of the combined flow. Attempting to model this effect accurately
would be impossible without making explicit assumptions for the
form of the distribution function. A similar difficulty may be involved with
the incompletely modeled radial motions of the saucer orbits, which are
very sensitive both to the start space coordinates and the shape of the
potential. But since it is dubious, at best, whether either of these effects
could ever be discerned observationally, there is little benefit in
overly-precise modeling.

A far more important effect that no one, including us, has yet dealt with
properly is figure rotation. We have repeated a fraction of the orbit survey,
rotating the potential about its short axis.
At dimensionless pattern speeds $\Omega$
on the order of a few $10^{-2}$ or less we find that the S tube streamlines
are mostly undisturbed, for both prograde and retrograde orbits.
This would correspond to locations in the galaxy on the order of 1/10
of the corotation radius, or to figure rotation periods on the order
of 10 dynamical times. The O tubes and box (as well as boxlet) orbits,
however, are distorted in more complicated ways (de Zeeuw \& Merritt
\markcite{dZM83}1983). Regular box orbits acquire a net prograde circulation
in the rotating frame; we find a rather box-like mean flow around the
perimeter of the orbit envelope. O tubes have their symmetry with respect to
the $yz$ plane broken, and averaged prograde-retrograde pairs show a complex
circulation about the $z$ axis. Neither of these effects can be approximated
by confocal streamlines.
More sophisticated methods will need to be developed to accurately model
triaxial systems with rotating figures.

\acknowledgments

This work was supported by NASA Astrophysical Theory Program grants
NAG5-2860 and NAG5-3050. We thank Suvendra Dutta for helpful comments
on the manuscript.

\begin{deluxetable}{cccccc}
\tablewidth{0pc}
\tablecaption{Model Shape Parameters \label{t:models}}
\tablehead{\colhead{\null} & \multicolumn{2}{c}{Density} &
\multicolumn{2}{c}{Potential} & \multicolumn{1}{c}{Streamlines} \\
\colhead{Model} & \colhead{$T_{m}$} & \colhead{$c_m$} &
\colhead{$T_p$} & \colhead{$c_p$} &
\colhead{$\langle T_{\rm fit}\rangle$}}
\startdata
1 & 0.50 & 0.7 & 0.472 & 0.889 & 0.480 \nl
2 & 0.50 & 0.5 & 0.467 & 0.799 & 0.473 \nl
3 & 0.50 & 0.3 & 0.479 & 0.688 & 0.508 \nl
4 & 0.10 & 0.3 & 0.096 & 0.712 & 0.111 \nl
5 & 0.90 & 0.3 & 0.883 & 0.649 & 0.898 \nl
6 & 0.98 & 0.3 & 0.976 & 0.630 & 0.969 \nl
\enddata
\end{deluxetable}

\begin{deluxetable}{rccccr}
\tablewidth{0pc}
\tablecaption{Averaged Streamline Parameters \label{t:averages}}
\tablehead{\colhead{Orbits} & \colhead{$\langle T \rangle$} &
\colhead{$\sigma_T$} & \colhead{$\langle \Delta \rangle$} &
\colhead{$\sigma_\Delta$} & \colhead{$N$}}
\startdata
Model 1: O & 0.484 & 0.010 & 0.223 & 0.067 & 26 \nl
         S & 0.476 & 0.010 & 0.208 & 0.073 & 27 \nl
       all & 0.480 & 0.011 & 0.215 & 0.071 & 53 \nl
Model 2: O & 0.477 & 0.011 & 0.315 & 0.057 & 16 \nl
         S & 0.464 & 0.026 & 0.280 & 0.070 & 28 \nl
       all & 0.473 & 0.032 & 0.297 & 0.081 & 46 \nl
Model 3: O & 0.523 & 0.034 & 0.415 & 0.035 & 20 \nl
         S & 0.499 & 0.079 & 0.371 & 0.058 & 20 \nl
         s & 0.497 & 0.058 & 0.344 & 0.066 & 11 \nl
       all & 0.508 & 0.060 & 0.381 & 0.058 & 53 \nl
Model 4: O & 0.112 & 0.009 & 0.337 & 0.046 & 10 \nl
         S & 0.097 & 0.041 & 0.346 & 0.102 & 28 \nl
         s & 0.163 & 0.037 & 0.297 & 0.044 &  8 \nl
       all & 0.111 & 0.043 & 0.335 & 0.087 & 46 \nl
Model 5: O & 0.906 & 0.018 & 0.507 & 0.075 & 17 \nl
         I & 0.908 & 0.007 & 0.483 & 0.034 &  6 \nl
         S & 0.877 & 0.004 & 0.380 & 0.032 & 14 \nl
       all & 0.898 & 0.024 & 0.459 & 0.084 & 38 \nl
Model 6: O & 0.970 & 0.021 & 0.434 & 0.153 & 20 \nl
         I & 0.970 & 0.019 & 0.425 & 0.126 &  9 \nl
         S & 0.965 & 0.004 & 0.356 & 0.044 &  8 \nl
       all & 0.969 & 0.018 & 0.415 & 0.134 & 37 \nl
\enddata
\end{deluxetable}


\newpage
\begin{references}
\reference{Arn94} Arnold, R., de Zeeuw, P. T., \& Hunter, C. 1994, \mnras,
	271, 924
\reference{Bin85} Binney, J. 1985, \mnras, 212, 767
\reference{Bin87} Binney, J. 1987, in Structure and Dynamics of
	Elliptical Galaxies, T. de Zeeuw, ed., (Dordrecht: Reidel), p. 229
\reference{BandT} Binney, J. \& Tremaine, S. 1987, Galactic Dynamics
	(Princeton: Princeton Univ. Press)
\reference{dZM83} de Zeeuw, T. \& Merritt, D. 1983, \apj, 267, 571
\reference{deZ85} de Zeeuw, T. 1985, \mnras, 216, 273
\reference{Fra91} Franx, M., Illingworth, G. \& de Zeeuw, T. 1991, \apj,
	383, 112
\reference{Lau95} Lauer, T. R., Ajhar, E. A., Byun, Y.-I., Dressler, A.,
	Faber, S. M., Grillmair, C., Kormendy, J., Richstone, D., \& Tremaine,
	S. 1995, \aj, 110, 2622
\reference{Lee92} Lees, J. \& Schwarzschild, M. 1992, \apj, 384, 491
\reference{Lyn62} Lynden-Bell, D. 1962, \mnras, 124, 95
\reference{MeF96} Merritt, D. \& Fridman, T. 1996, \apj, 460, 136
\reference{MeV96} Merritt, D. \& Valluri, M. 1996, \apj, 471, 82
\reference{Mir89} Miralda-Escud\'e, J. \& Schwarzschild, M. 1989, \apj, 339, 752
\reference{Sch93} Schwarzschild, M. 1993, \apj, 409, 563
\reference{Sta91} Statler, T. 1991, \aj, 102, 882 
\reference{Sta94a} Statler, T. 1994a, \apj, 425, 458
\reference{Sta94b} Statler, T. 1994b, \apj, 425, 500
\reference{Sta96fv} Statler, T. 1996, in Fresh Views of Elliptical
	Galaxies, A. Buzzoni, A. Renzini, \& A. Serrano, eds.
	(San Francisco: Astronomical Society of the Pacific), p. 27
\reference{Ten92} Tenjes, P. , Busarello, G., \& Longo, G. 1992, Triaxial
	Elliptical Galaxies With Dust Lanes: An Atlas of Velocity Fields,
	(Napoli: Liguori editore).
\reference{Ten93} Tenjes, P., Busarello, G., Longo, G., \& Zaggia S. 1993,
	\aap, 275, 61

\end{references}
\end{document}